\documentclass[preprint,showpacs,preprintnumbers,amsmath,amssymb]{revtex4}
\usepackage{graphicx}
\begin{document}
\title{Model Dependence of the Properties of $S_{11}$ Baryon Resonances}
\preprint{nucl-th/0308066}
\author{Alvin Kiswandhi}
\email{alvin@heisenberg.physics.fsu.edu}
\affiliation{
  Department of Physics, Florida State University,
  Tallahassee, FL 32306-4350, U.S.A.}
\author{Simon Capstick}
\email{capstick@phy.fsu.edu}
\affiliation{
  Department of Physics, Florida State University,
  Tallahassee, FL 32306-4350, U.S.A.}
\author{Steven Dytman}
\email{dytman+@pitt.edu}
\affiliation{
  Department of Physics and Astronomy, University of Pittsburgh,
  3941 O'Hara St., Pittsburgh, PA 15260, U.S.A.}
\date{\today}
\begin{abstract}
The properties of baryon resonances are extracted from a complicated
process of fitting sophisticated, empirical models to data.  The
reliability of this process comes from the quality of data and the
robustness of the models employed.  With the large of amount of data
coming from recent experiments, this is an excellent time for a study
of the model dependence of this extraction process.  A test case is
chosen where many theoretical details of the model are required, the
$S_{11}$ partial wave.  The properties of the two lowest $N^*$
resonances in this partial wave are determined using various models of
the resonant and non-resonant amplitudes.
\end{abstract}
\maketitle
%
%
\def\lpmb#1{\mbox{\boldmath $#1$}}
\newcommand{\Meff}{M_{\mbox{\small eff}}}
\newcommand{\vbead}{V_{\mbox{\small J}}}
\newcommand{\bp}{{\bf p}}
\newcommand{\bk}{{\bf k}}
\newcommand{\bx}{{\bf x}}
\newcommand{\by}{{\bf y}}
\newcommand{\bz}{{\bf z}}
\newcommand{\btab}{\begin{tabbing}}
\newcommand{\etab}{\end{tabbing}}
\newcommand{\eqntimes}{\mbox{} \times}
\newcommand{\eqnhspace}{\hspace{3em}}
\newcommand{\intereqnvspace}{\vspace{-2.1ex}}
\newcommand{\eqnneghspace}{\hspace{-6ex}}
\newcommand{\beqn}{\begin{equation}}
\newcommand{\eeqn}{\end{equation}}
\newcommand{\barr}[1]{\begin{array}{#1}}
\newcommand{\earr}{\end{array}}
\newcommand{\beqna}{\begin{eqnarray}}
\newcommand{\eeqna}{\end{eqnarray}}
\newcommand{\btablec}{\begin{table} \begin{center}}
\newcommand{\etablec}{\end{center} \end{table}}
\newcommand{\lapprox}{\stackrel{<}{\scriptstyle \sim}}
\newcommand{\rapprox}{\stackrel{>}{\scriptstyle \sim}}
\newcommand{\gapproxeq}{\lower.7ex\hbox{$\;\stackrel{\textstyle>}
{\sim}\;$}}

\sloppy
\bigskip

\section{Introduction}

The problem of extracting baryon resonance parameters from observables
measured in scattering experiments is of fundamental importance. In
order to understand the relevant degrees of freedom at low energy and
their interactions in the few dozen $N^*$ states found to date,
reliable and objective information on their properties extracted from
data is required. Models of the baryon spectrum are usually compared
to masses defined in terms of effective Breit-Wigner parameters. A
more sensitive test of such models is the comparison of total widths
and branching fractions to open channels of each resonance to those
extracted from data. These quantities are also defined in terms of
effective Breit-Wigner parameters, but in general show more
sensitivity to the method used to fit the data.  What is not commonly
appreciated is the dependence of these resonance parameters on the
method used to extract them from a fit to the data. For example, it is
not uncommon to see such parameters referred to as ``data'' in the
literature.

The primary goal of this paper is to examine the model dependence in
this process in a carefully chosen test case, states in the $S_{11}$
partial wave in $\pi N$.  States in this partial wave have both
significant physics interest and significant uncertainty in their
parameters as reported by the Particle Data Group (PDG) in their
Review of Particle Properties~\cite{PDG}.  Most of the literature
considers the lowest energy resonance in this partial wave
[$S_{11}$(1535)] as a three-quark state within the quark
model~\cite{quarkmodels}. However, the substantial branching fraction
of $S_{11}$(1535) to $\eta N$ has, given the small phase space
available for this decay, led to the interpretation of this state as a
meson-baryon molecule~\cite{chiral}. Lattice calculations of the
masses of these states have recently been
developed~\cite{lattice1,lattice2}.  The calculations of both
Ref.~\cite{lattice1} and~\cite{lattice2} are quenched, and find masses
for $S_{11}$(1535) similar to the values fit from
experiment~\cite{PDG}.  In addition, constituent quark models which
describe these $S_{11}$ resonances as ($P$-wave) orbital excitations
with $J^P=1/2^-$ differ in their prediction of the amount of mixing
between the two possible quark-spin states~\cite{Karl}, because of
different models of the short distance interactions between the
quarks. A comparison between partial widths calculated in a given
model and those extracted from data can be used to determine the
nature of these short distance interactions. This makes the dependence
of these partial widths on the method used to extract them from fits
to data important, and relevant.

These important microscopic issues are unable to be settled with the
uncertainty in the full and partial widths presently reported by the
PDG, which vary widely among various studies.  At one extreme,
analyses of the $\pi N$ elastic scattering data tend to give a full
width of about 120 MeV for $S_{11}$(1535), with $\pi N$ as the
dominant decay branch~\cite{VPIpiN,KH80}.  At the other extreme,
threshold eta photo-production data gives a much larger full width
$\Gamma \sim$ 210 MeV, with $\eta N$ as the larger decay 
branch~\cite{KruMuk}. Only a unified treatment of all the data can 
provide a consistent picture.

The purpose of this paper is to examine the model dependence of
$S_{11}$ resonance parameters by extracting them from several fits to
a single set of
partial wave amplitudes for $\pi N\to \pi N$~\cite{VPIpiN} and
$\pi N\to \eta N$~\cite{Vrana} scattering in the $S_{11}$ partial
wave.  These fits have different levels of sophistication in their of
description of the scattering matrix. The analysis of these channels
in this partial wave is chosen because the proximity of two
overlapping resonances [$S_{11}(1535)$ and $S_{11}(1650)$] and the
$\eta N$ channel threshold makes it an interesting and non-trivial
example.  Differing treatments of the re-scattering processes are
a key difference among the models studied here and these effects 
should be most important near a threshold.  Although the number of 
open channels is much larger than
two, the two channels that have been chosen are the most important and
account for 90\% (1535) and about 80\% (1650) of the decay width when
all channels are included~\cite{PDG}.

Breit-Wigner models can get good fits to single channel data 
(e.g. $\pi N \to \eta N$~\cite{Baker}, $\gamma N \to \pi N$~\cite{MAID}), 
but have non-trivial
uncertainties and at times surprising results.  In simplifying an 
intrinsically multichannel problem
into a single channel model, information from other reactions must be
included with accompanying uncertainties.  In addition, there is no
single commonly accepted form for the resonance shape.  
When a somewhat large ($\sim$10) set of asymptotic states can 
couple to each $N^*$ with high probability, including a large
sample of reactions and application of the unitarity constraint 
are extremely important.  Common
multichannel techniques include $K$-matrix~\cite{Penner,Green} and
Carnegie-Mellon Berkeley (CMB)~\cite{Batinic,Vrana} models.  Each 
has an established way of including a
variety of interactions respecting multichannel unitarity with
varying ability to include dynamics.

Three methods are used here to perform the extraction of resonance
parameters via a fit to the partial wave amplitudes. The most
constrained is the CMB model, which was developed~\cite{Cutkosky} to
describe $\pi N$ elastic and inelastic scattering, and extended and
modernized in Ref.~\cite{Vrana}.  This approach has multi-channel
unitarity, and the scattering matrix (a matrix in the space of
channels) has the required analytic structure.  Resonances are modeled
as ``bare'' poles, which are ``dressed'' by coupling to the open
asymptotic channels.

The second, simpler approach is to describe the unitary $T$ matrix
using a real symmetric $K$ matrix, which is in turn written directly
in terms of resonance parameters. This approach maintains
multi-channel unitarity, but does not satisfy analyticity
constraints. The third, simplest approach is to build the $T$ matrix
directly from a sum of relativistic Breit-Wigner forms for the
resonances.  In this case, minimal constraints on the scattering
amplitudes are available.  Below $\pi \pi$ production threshold, the
overall amplitude can be made unitary using Watson's theorem.
However, the $\eta N$ channel must be included for a good description
of the $S_{11}$ partial wave, and only ad-hoc methods~\cite{MAID} are
available to accomplish this.  The Breit-Wigner models used here are
neither unitary nor analytic.

Non-resonant amplitudes must be added to account for scattering
processes which do not involve $s$-channel $N^*$ resonances.  Although
a fairly small set of diagrams describing non-resonant processes can
be identified at lower total energy ($W\sim$ 1.3 GeV), the set of
possible diagrams grows rapidly as the total energy increases to 2.0
GeV.  To date, no published work has included all the relevant
diagrams.  Empirical descriptions of the non-resonant amplitudes can
be chosen since each resonance has a strong signal in at least one of
the reactions studied, and most publications find the non-resonant
amplitudes to be smaller than the resonant amplitudes.  In addition,
the most basic physics assumption is that the resonances come from the
long-distance part of the interaction and are seen in the sharp energy
features of the data, while the non-resonant amplitudes arise from the
short-range part of the interaction and provide smooth energy
dependence.  This implies that, in this case, the influence of the
choice of non-resonant amplitudes on the extracted resonance
properties should be small. The ``distant poles'' model of the
non-resonant amplitudes is designed for use with the CMB model.  Bare
poles well below and well above the thresholds for the channels being
studied, i.e. distant poles, are fit to the partial-wave data with
methods very similar to those used for the resonance poles. In the CMB
model they provide a smooth, analytic, unitary $T$ matrix in the
physical region; they can also be incorporated into a $K$-matrix model
to provide a unitary scattering matrix.  While the poles and cuts in
the left-hand part of the complex $s$-plane required by crossing
symmetry are difficult to fully implement in the CMB model, their
effects in the physical region can be simulated by a particular choice
of strong form factor, and by the use of sub-threshold poles. Another
method adds to the resonant amplitudes polynomials quadratic in the
parameters $x_c=(s-s_{{\rm th},c})^{1/2}$ for each exit channel $c$,
which can yield a unitary $T$ matrix for the $K$-matrix approach.

The procedure used here is to calculate, using the CMB, $K$-matrix,
and Breit-Wigner approaches, the matrix $T_{ab}$ for scattering
between the channels $\{a,b\}\in\{\pi N,\eta N\}$ in terms of a set
of parameters describing the resonant and non-resonant
amplitudes. Different treatments of the background are explored, giving
a total of eight different models. A $\chi^2$
statistic is minimized by varying these parameters to best fit the
partial wave amplitudes for the $\pi N \to \pi N$~\cite{VPIpiN}
and $\pi N \to \eta N$~\cite{Vrana} reactions.

The goals of this work are somewhat limited in order to make a clear
statement about model dependence, and for this reason the resonance
parameters which result should not be considered for use in
understanding the microscopic structure of the $S_{11}$ resonance
states.  The channel space is greatly simplified to only the two
principal channels, leaving out other channels such as $\pi \Delta$
that are a much smaller part of the amplitudes, but which must be
included for the optimal values.  From ten-channel fits to both $\pi
N$ and $\gamma N$ reactions, the $S_{11}$(1650) width is known to be
$\sim$50\% larger than in a two-channel fit.  In addition, the
sophisticated $K$-matrix techniques developed by the Giessen
group~\cite{Penner} are not used here.

\section{Models of the scattering matrix\label{Models}}

In this section the form of the scattering matrix in the various
models is described, along with the implementation of the two forms of
background described above.  All of the models used here require
partial wave amplitudes (PWA) as input.  These extract the energy
dependence of amplitudes with specific isospin, parity and angular
momentum (e.g. $S_{11}$) from the large number of experimental data
points.  About 20 years ago, significant efforts were made to find
model independent methods for partial wave analysis of $\pi N$ elastic
data~\cite{KH80}.  More recent efforts by the George Washington
University (GWU) group~\cite{VPIpiN}
incorporate most of the theoretical constraints developed previously.
Although the fits in this work do not use the most recent GWU work,
results using the most recent PWA would not produce different
conclusions.  Since the number of data points for the $\pi N
\rightarrow \eta N$ reaction is much smaller, there is more model
dependence.  These amplitudes are discussed in Ref.~\cite{Vrana},
using a method that that accurately couples the $\pi N$ elastic and
inelastic data with minimal model dependence.  Until another study is
done, this is the only PWA result available for $\pi N \rightarrow
\eta N$.

\subsection{CMB model}

In this work the $S_{11}$ partial wave in $\pi N$ elastic and $\pi
N\to \eta N$ scattering is described using only the two light
resonances $S_{11}$(1535) and $S_{11}$(1650). In the CMB model
background ($t$- and $u$-channel) processes are simulated using one
high-energy and two sub-threshold $s$-channel poles, which are treated
identically to the resonances in order to preserve the analytic
structure of the $T$ matrix.  The transition
amplitudes~\cite{Cutkosky,Vrana} are
\beqn
T_{ab}=\sum_{i=1}^5\sum_{j=1}^5
f_a(s) \sqrt{\rho_a(s)} \gamma_{ai} G_{ij}(s)
\gamma_{bk} \sqrt{\rho_b(s)} f_b(s)
\label{Tmatrix}
\eeqn
where $\{a,b\}\in\{\pi N,\eta N\}$ label the channels, and $i,j$ label
the poles, which can represent the resonances or the background. 
The model used in this paper has five poles (two for real
resonances and three for background resonances).  It is easy to add 
channels with this formalism if the input data is available.
The phase space factor
$\rho_a(s)$ has the form
\beqn
\rho_a(s)=\frac{p_a}{\sqrt{s}},
\label{rho}
\eeqn
where $p_a$ is the c.m. frame momentum and $\gamma_{ai}$ is the
real-valued coupling constant of resonance $i$ to the channel $a$.
The form factor $f_a$ in Eq.~(\ref{Tmatrix}) is unity in the $S$-wave.

The dressed propagator $G_{ij}(s)$ allows resonance $i$ to couple to
resonance $j$ through re-scattering, and is the solution of the Dyson
equation
\beqn
G_{ij}(s)=G^0_{ij}(s)+G^0_{ik}(s)\Sigma_{kl}(s)G_{lj}(s).
\label{Gij}
\eeqn
The bare propagator $G^0_{ij}(s)$ has the form
\beqn
G^0_{ij}(s) = {\delta_{ij}e_i \over s_{0i}-s},
\eeqn
where $e_i=+1$ for the resonances and the high-energy background pole,
and $s_{0i}$ is the bare mass of the resonance (background pole). One
sub-threshold pole has $e_i=+1$ (repulsive), and the other has
$e_i=-1$ (attractive). 

The self energy $\Sigma_{kl}$ in Eq.~(\ref{Gij}) is the sum over
channels $c$
\beqn
\Sigma_{kl} = \sum_{c=1}^2\gamma_{ck}\phi_c(s)\gamma_{cl},
\label{Sigma}
\eeqn 
where $\phi_c(s)$ is the channel propagator, which plays a central
role in the CMB model. The sum is over the two channels in this
simplified problem.  Two-pion channels such as $\rho N$ can be
included by treating them as quasi-two-body channels, which results in
an enlarged $T$ matrix. The re-scattering sum in Eq.~(\ref{Sigma}) will then
extend over the additional channels. The imaginary part of $\phi_c$ is
the product of phase space, $\rho_c$(s), with the square of the form
factor for channel $c$ (this quantity will be seen in the other models
below) and the real part is obtained through a once-subtracted
dispersion integral~\cite{Vrana,Cutkosky}.  The desired analyticity of
the entire amplitude is then assured.

Unitarity comes about because of the democracy of the re-scattering
in the Dyson equation, Eq.~(\ref{Gij}).  In the multichannel
context, unitarity is expressed through a generalized optical
theorem,
\beqn
{\rm Im} T_{ab}= \sum_c{T_{ac}^*T_{cb}}
\eeqn
Unitarity is also related to the properties in the complex
plane through the discontinuity in the amplitude across
the right-hand cut.

The dressed propagator $G_{ij}(s)$ can be found by solving
algebraically the matrix equation Eq.~(\ref{Gij}), with the result
\beqn
H_{ij}(s)\equiv [G^{-1}(s)]_{ij}=[G^0(s)^{-1}]_{ij}-\Sigma_{ij}(s)
={s_{0i}-s\over e_i }\delta_{ij}-\Sigma_{ij}(s),
\label{Hij(s)}
\eeqn
and the matrix $G(s)$ can be found by inverting $H(s)$. This completes
the necessary ingredients for calculating the $T$ matrix in the CMB
model of Eq.~(\ref{Tmatrix}) in terms of three parameters for each
resonance or background pole. These are the bare mass squared
$s_{0,i}$, and the coupling strengths $\gamma_{\pi N,i}$ and
$\gamma_{\eta N,i}$. Since there are five poles, this model has a
total of fifteen parameters.

Once this $T$ matrix is fitted to the partial wave data by varying
these fifteen parameters, baryon resonance parameters are extracted
from the resonant part of the $T$ matrix. The procedure for doing this
is described in what follows.

\subsection{Extraction of resonance parameters in the CMB model}
Poles in the resonant part of the $T$ matrix occur at complex values
of $s$ where the denominator of the resonance propagator vanishes.  A
search program finds these pole locations, and these are the model-free
output (see Ref.~\cite{Vrana} for
details). This is possible because the amplitude has reasonable
properties for complex values of $s$.

However, it is important to derive effective Breit-Wigner parameters
for each resonance from this model.  Once the pole positions are
found, the matrix $H(s_{\rm pole})$ is diagonalized to eliminate
interference between resonances. The denominator of the resonant 
part of the $T$ matrix
in the vicinity of each pole is then expanded in $s$.  The constant
term becomes the dispersive correction to the mass, and the term
linear in $s$ becomes the width of a generalized relativistic
Breit-Wigner form.  Thus, although the energy dependence of a
resonance in the CMB model is much more sophisticated than the normal
Breit-Wigner shape, all characteristics of a resonance can be
expressed in this commonly used form.

\subsection{CMB model with polynomial background}

This is a modified CMB model which uses a polynomial function to
describe the background. Care must be taken to maintain unitarity of
the $T$ matrix. Using a technique from Ref.~\cite{Manley}, the full
$S$ matrix is written
\beqn S=B^\dag RB,
\label{SCMB-poly}
\eeqn 
where $R$ and $B$ are the resonant and non-resonant $S$ matrices,
respectively.  Here $R=I+2iT$ uses the CMB-model $T$ matrix, and $B$
is a unitary matrix that need not be symmetric. The background matrix
$B$ is constructed from a real symmetric matrix $K^B$
\beqn
B=\left(1+iK^B\right)\left(1-iK^B\right)^{-1},
\eeqn
where for the two channel case of interest 
\beqn
K^B=\left[ \barr {cc} k_{\pi\pi} x_\pi + k^\prime_{\pi\pi} x_\pi^2
& k_{\pi\eta} x_\eta + k^\prime_{\pi\eta} x_\eta^2 \\ 
k_{\pi\eta} x_\eta + k^\prime_{\pi\eta} x_\eta^2  & 
k_{\eta\eta} x_\eta +k^\prime _{\eta\eta} x_\eta^2 \\ \earr \right],
\label{Kbgrd}
\eeqn
or 
\beqn
K^B_{ab}=k_{ab} x_b + k^\prime_{ab} x_b^2,
\eeqn
$x_b=\sqrt{s-\left(m_N+m_b\right)^2}$ for $s$ above exit channel
threshold and zero otherwise, and the coefficients $k_{ab}$ and
$k^\prime_{ab}$ are real. Note that the off-diagonal terms in $K^B$
are zero for $s$ below $\eta N$ threshold. This ensures that the
diagonal and off-diagonal elements of $R$ are not mixed, so 
the $\pi N \to \eta N$ amplitude vanishes below $\eta N$
threshold.  This model has six real parameters describing the
background, and six describing the resonances, for a total of twelve
real parameters.

Once the $S$ matrix in Eq.~(\ref{SCMB-poly}) is formed, it can be
easily converted using $S=I+2iT$ to a $T$ matrix that can be fit to
the partial wave amplitudes by varying these twelve parameters in the
usual way, and the extraction of the baryon parameters is identical to
that of the original CMB model.

\subsection{K-Matrix models}
A unitary $S$ matrix can be constructed from a real, symmetric
$K$-matrix via $T=K(1-iK)^{-1}$.  This method is very common in the
literature~\cite{Longacre,Moorhouse,Penner,Chung} for investigating
hadrons in reactions.  Recently, effective Lagrangian
models~\cite{Penner} have been used with a $K$-matrix for studies of
$N^*$ states.  Non-relativistic reductions of Feynman diagrams make
these studies more sophisticated than what is presented here.  The
earlier work of Moorhouse and collaborators~\cite{Moorhouse} is very
similar to the method used here.  

The differences between the CMB and $K$-matrix models are well known.
A common problem with all $K$-matrix methods is the difficulty in
maintaining analyticity.  Only the work of Longacre~\cite{Longacre}
accomplished this.  Thus, pole positions and residues can be obtained
from the CMB fits and not from the $K$-matrix fits.  Many researchers
feel these values have less model dependence.  Any
microscopic model (e.g. lattice QCD) which has the analytic properties as
an output can only use the CMB results.  While the CMB model has a
full treatment of re-scattering, only the imaginary part of the channel
propagator is included in the $K$-matrix model. The choice of the
$S_{11}$ resonances near $\eta N$ threshold as a test case is
motivated by the expectation that these re-scattering effects are most
important close to threshold.  As discussed above, the $K$-matrix
method can be incorporated directly into an effective Lagrangian
formulation~\cite{Penner}.  As a result, the non-resonant amplitudes
can be added diagram by diagram.  This is a significant advantage over
the present version of the CMB model, which is formulated in terms of
amplitudes rather diagrams.  In this work, we use the amplitude form
of the $K$-matrix model~\cite{Moorhouse, Chung}.

Although the CMB model is preferred on theoretical grounds, the
$K$-matrix model has practical advantages.  It is more commonly used,
and the less complicated treatment of interference and re-scattering
effects simplifies the fits to partial-wave amplitudes.

As in the CMB model, two sub-threshold poles and one high-energy pole
can be used to describe the non-resonant part of the scattering
matrix. The contributions from the resonance and non-resonant poles
are
\beqn
K_{ab}^i(s)={\sqrt{\rho_a(s)} f_a(s)
\gamma_{ai}\gamma_{bi} f_b(s) \sqrt{\rho_b(s)}
\over\left(s_i-s\right)e_i}.
\label{kiab}
\eeqn 
Here the form is similar on the surface to what is used in CMB model.
The phase space $\rho_a(s)$ is given by Eq.~(\ref{rho}), $f_a(s)$ is
the form factor (none is needed for $S$-wave), $\gamma_{ai}$ is the
coupling constant of the resonance or background pole $i$ to channel
$a$ (but with units different from the CMB model), and $s_i$ is the
position of the $i$-th (real) pole. The signs $e_i$ are the same as
those used in CMB model, with $e=+1$ for the resonances, the
high-energy background pole, and one sub-threshold background pole,
while the second sub-threshold background pole has $e=-1$. The
$K$-matrix is formed from a simple sum over the resonance and
background poles
\beqn
K_{ab}(s)=\sum_{i\in {\rm poles}} K_{ab}^i(s)
\eeqn

When the $K$-matrix is converted to a $T$-matrix, re-scattering
(``dressing'' of the resonance) comes about naturally and the
resonance gains a finite width.  In this model the $K$-matrix is
written directly in terms of the resonance parameters, so that
$M_i=\sqrt{s_i}$ is the mass of the resonance associated with the pole
at $s_i$, the partial width of the $i$-th resonance to channel $a$ is
\beqn 
\Gamma_{ai}={\rho_a(s_i) f_a^2 \gamma_{ai}^2 \over M_i},
\label{gamma-ai}
\eeqn
and the total width is the sum of the partial widths.  As discussed
above, the differences between CMB and $K$-matrix formalism are seen
only when interactions with other resonances and non-resonant amplitudes
occur. These differences are expected to be maximized close to a
channel threshold, which is the reason for the choice of the $S_{11}$
partial wave with resonances near $\eta N$ threshold. In this case,
the CMB model is expected to give more reliable results.

An alternate description of the background is using the polynomial
form of Eq.~(\ref{Kbgrd}), which is added to the resonance $K$ matrix
\beqn
K_{ab}=\sum_{i\in {\rm resonances}} K_{ab}^i(s)+K^B_{ab}.
\eeqn
With either method of treating the background, these $K$-matrix models
have the same total number of parameters as the CMB models, 15 for the
distant-poles background and 12 for the polynomial background.

\subsection{Breit-Wigner models}
There are two kinds of Breit-Wigner model used here along with two
kinds of non-resonant amplitude parameterization. In each model,
the resonant and non-resonant $T$-matrices are summed,
\beqn
T_{ab}(s)=\sum_{i\in {\rm resonances}}T_{ab,{\rm res}}^i(s)
+T_{ab,{\rm nonres}}(s),
\label{BW-gen}
\eeqn
This form is inconsistent with unitarity without special effort.

The resonant amplitude $T_{ab,res}^i(s)$ is given by
\beqn
T_{ab,{\rm res}}^i(s)={\sqrt{\Gamma_{ai}\Gamma_{bi}}\sqrt{s}
\over e_i\left(s_i-s\right)-i\sqrt{s}\Gamma_i}e^{i\theta^{ab}_i}.
\label{BW}
\eeqn
The signs $e_i$ are the same as in the CMB and $K$-matrix models,
$s_i$ is the pole position (with $M_i=\sqrt{s_i}$ for resonance
poles), and $\Gamma_i=\sum_a\Gamma_{ai}$ is the total width of
resonance $i$. Using a procedure similar to that of Ref.~\cite{MAID}, an
arbitrary phase $\exp(i\theta^{ab}_i)$ is associated with each pole in each
channel. As shown below, the freedom to fit these additional phases is
crucial to achieving a good fit with a Breit-Wigner model to the
$S_{11}$ partial-wave $T$ matrix elements in these channels.

The width commonly used to describe a non-relativistic Breit-Wigner
form is the energy-dependent width
\beqn
\Gamma^{\rm n.r.}_{ai}(s)=\Gamma_{ai}(s_i){\rho_a(s)\over\rho_a(s_i)},
\eeqn
with the phase-space factor $\rho_a(s)$ as in Eq.~(\ref{rho}).  
(Note that this is not the same as the conventional non-relativistic
Breit-Wigner energy dependence.)
An alternative, relativistic form results from the
assumption that the numerator of the contribution of resonance $i$ to
the $K$-matrix in Eq.~(\ref{kiab}) is the same as that in the
Breit-Wigner form in Eq.~(\ref{BW}). If the partial width used in the
$K$-matrix formalism, Eq.~(\ref{gamma-ai}), is generalized to an
energy-dependent function by replacing $M_i$ with $\sqrt{s}$, then the
$K$-matrix numerator of Eq.~(\ref{kiab}) is
\beqn
\sqrt{\rho_a(s)} f_a(s) \gamma_{ai}\gamma_{bi}
\sqrt{\rho_b(s)} f_b(s)
=\sqrt{\Gamma_{ai}\sqrt{s}}\sqrt{\Gamma_{bi}\sqrt{s}}
=\sqrt{\Gamma_{ai}\Gamma_{bi}}\sqrt{s},
\eeqn
which is the same as the numerator in Eq.~(\ref{BW}). This suggests
the use of
\beqn
\Gamma^{\rm rel}_{ai}(s)=\Gamma_{ai}(s_i){\sqrt{s_i}\over \sqrt{s}}
{\rho_a(s)\over\rho_a(s_i)},
\eeqn
for the energy-dependent partial width in a relativistic Breit-Wigner
form. In what follows models using both $\Gamma^{\rm n.r.}_{ai}(s)$,
labeled BW$_{\rm n.r.}$, and $\Gamma^{\rm rel}_{ai}(s)$, labeled
BW$_{\rm rel}$, are used in fitting the Breit-Wigner form in
Eq.~(\ref{BW}) to the $S_{11}$ partial-wave $T$ matrix elements, and
the results are compared. The relativistic form is the same as that
advocated by Chung et al.~\cite{Chung}.

The first non-resonant form uses
distant poles to describe the background contributions
to the $T$ matrix, so that the total $T$-matrix is a simple sum
\beqn
T_{ab}(s)=\sum_{i\in {\rm poles}} T_{ab}^i(s),
\eeqn
This form of background has two low energy poles and one high energy
pole, the same as was used with the CMB and $K$-matrix models.  To get
a good fit, each of these terms requires an additional phase in each
channel, as described above.  With the background described by distant
poles, there are 15 parameters (mass, width, and $\pi N$ branching
fraction for each term) associated with the five poles, and ten
phases, for a total of 25 real parameters. Of the models used in this
work, this model has the largest number of parameters.

As with CMB and $K$-matrix models, it is possible to describe the 
non-resonant background by a polynomial function, so that
\beqn
T_{ab}(s)=\sum_{i\in {\rm resonances}}T_{ab,{\rm res}}^i(s)
+T_{ab,{\rm nonres}}(s),
\eeqn
where the background $T$-matrix elements are the polynomials
\beqn
T_{ab,{\rm nonres}}=\kappa_{ab} x_b + \kappa^\prime_{ab} x_b^2.
\eeqn
As with the other models, $x_b=\sqrt{s-\left(m_N+m_b\right)^2}$ for
$s$ above exit channel threshold and zero otherwise, but now the
coefficients $\kappa_{ab}$ and $\kappa^\prime_{ab}$ are complex. In
this model there are six resonance parameters associated with the two
poles, and 8 background parameters (2 complex numbers for each of two
channels), for a total of 18 real parameters.

\section{Results}

\begin{figure}
\includegraphics[width=0.7\linewidth,angle=-90]{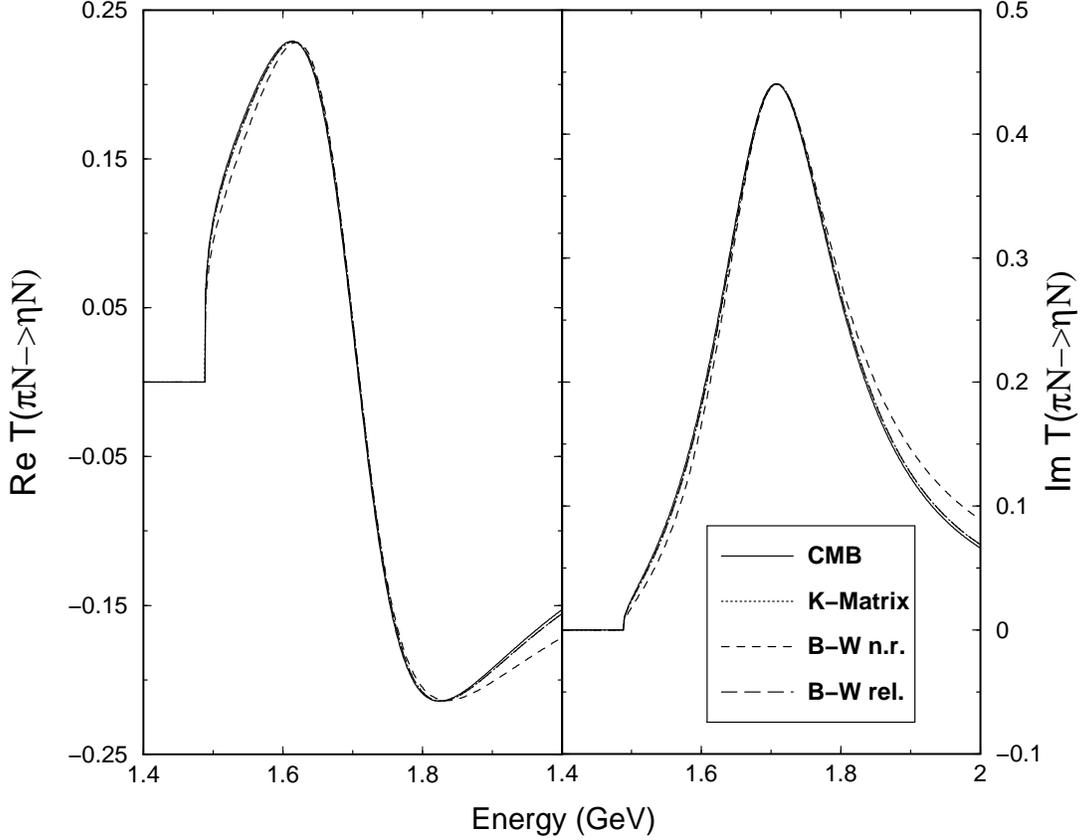}
\caption{\label{fig:singchan} Real parts (left panel) and imaginary parts
(right panel) of $\pi N\to\eta N$ resonant scattering amplitudes
calculated with an isolated $S$-wave resonance of mass 1710 MeV and
width 215 MeV, in each of the four models used
here.}
\end{figure}

In order to understand the complications in the $S_{11}$ partial wave,
model predictions for a hypothetical isolated resonance are first
considered.  Figure~\ref{fig:singchan} compares the results of all
four models for the scattering amplitude of an isolated single-channel
$S$-wave resonance. An $S$-wave resonance is chosen with a mass of 1710
MeV and width of 215 MeV in the $\pi N \to \eta N$ reaction, which
avoids any complications with centrifugal barriers. Agreement among
the different formulations is striking, despite the deliberate choice
of a resonance mass close to $ \eta N$ threshold.  The $K$-matrix and
CMB models have identical forms for an isolated resonance.  The
relativistic Breit-Wigner is chosen to be identical with the CMB
model in this limit.  Even the Breit-Wigner amplitude with the
non-relativistic width is in good agreement with the other other forms.
If the complex amplitudes are plotted on an Argand diagram (with the
real part of the amplitude on the horizontal axis, and the imaginary
part on the vertical axis), all will show the typical
counter-clockwise circular motion of a resonance.  Non-resonant
background will cause a shift or a distortion of this basic shape, and
inelasticity will decrease the radius.  Interference between
resonances can have significant effects.

\begin{figure}
\includegraphics[width=0.7\linewidth,angle=-90]{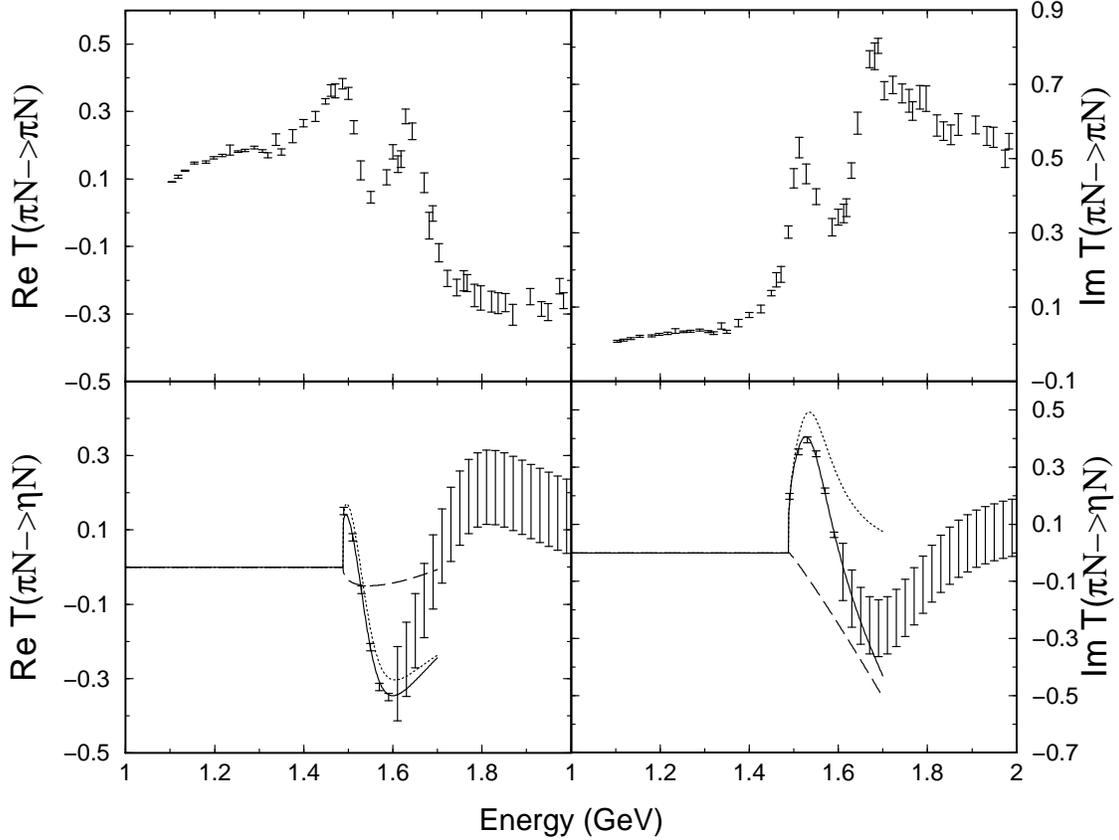}
\caption{\label{fig:data} Real parts (left panels) and imaginary parts
(right panels) of $\pi N\to\pi N$ (upper panels) and $\pi N\to\eta N$
(lower panels) scattering amplitudes, $T$, in the $S_{11}$ partial
wave. The curves in the lower panels are the single-channel,
one-resonance Breit-Wigner fit to these amplitudes described in the
text. Dotted lines give the resonant, dashed lines the non-resonant,
and solid lines the total scattering amplitude. Partial-wave
amplitudes are shown with error bars.}
\end{figure}

A partial wave amplitude with this $W$ dependence is the most visible
signature of a resonance, producing a peak in the total cross section
with an appropriate width.  Such a peak is seen in the total cross
section for $\pi N \to \eta N$, and is often interpreted as a
resonance.  The $S_{11}$ partial wave amplitude dominates the total
cross section for this reaction, and is shown in the bottom two panels
of Fig.~\ref{fig:data}.  Since the higher mass resonance
$S_{11}(1650)$ couples weakly to this channel, this appears to be an
isolated resonance with small non-resonant amplitudes.

Single-channel fits of the $\pi N\to \eta N$ amplitudes have been
made, which are shown as curves in the bottom two panels of
Fig.~\ref{fig:data}. The results are similar to those of
single-channel fits with a Breit-Wigner energy dependence to the total
cross sections for $\pi^- p\to \eta n$ and $\gamma p\to \eta
p$~\cite{KruMuk}.  The data can be fit using an $S$-wave Breit-Wigner
form if a small, but important contribution from non-resonant
amplitudes is included and the energy region used in the fit is
truncated.  This rough fit can be used to determine the product of the
initial and final state couplings to the resonance, requiring
information from more complete fits to account for the missing decay
strength of the $S_{11}$(1535) resonance to other channels. The
single-channel, one resonance fit shown in Fig.~\ref{fig:data} is made
with a Breit-Wigner resonant shape with the relativistic form of the
width, the polynomial form of the background, and a single phase
multiplying the resonant amplitude, truncating the fit at 1700
MeV. The result is a mass of 1543 MeV, a width of 134 MeV, and a $\pi
N$ branching fraction of 48\%.  This shows that it is possible to find
fit parameters that roughly agree with the more complete models, but
these results are of uncertain value given the strong model
dependence.

The elastic $\pi N$ amplitude (the
upper two panels of Fig.~\ref{fig:data}) is more complicated, showing
two structures overlapping in energy. The resonance $S_{11}(1650)$
couples strongly to $\pi N$ with peaks at $W\sim$ 1.7 GeV in Im($T$) 
{\it and} at $W\sim$ 1.65 GeV in Re($T$).  Although this rapid energy 
dependence is a clear resonant signal, there is also a
nontrivial non-resonant amplitude which greatly distorts the typical Argand
diagram.  The energy dependence of $S_{11}(1535)$ is more complicated
in $\pi N$ elastic scattering because of the strong
coupling to the $\eta N$ channel at its threshold ($W$=1.487 GeV),
which is within the resonant shape.  This produces a cusp in the real
part at threshold in addition to a peak in the imaginary part at
approximately the resonance mass.  Given this complicated structure,
analyticity of the scattering amplitude and dispersive scattering
effects can be expected to be most important for this state. For this
reason, the elastic amplitude includes many of the important dynamical
effects studied here.

\begin{figure}
\includegraphics[width=0.7\linewidth,angle=-90]{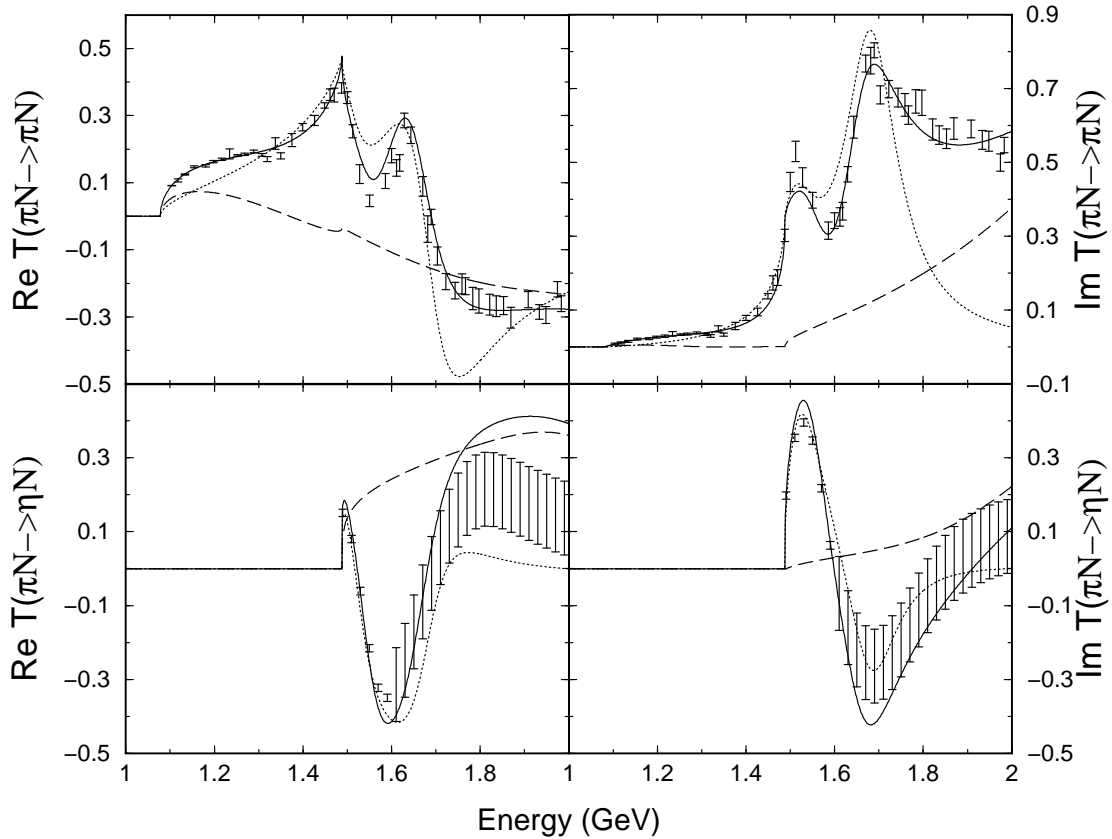}
\caption{\label{fig:cmb-distant} Real parts (left panels) and imaginary parts
(right panels) of $\pi N\to\pi N$ (upper panels) and $\pi N\to\eta N$
(lower panels) scattering amplitudes in the $S_{11}$ partial
wave. Dotted lines give the resonant, dashed lines the non-resonant,
and solid lines the total scattering amplitude. Partial-wave amplitudes 
are shown with error bars. The resonant model used here is the
two-resonance, two-channel CMB model described in the text, with
non-resonant amplitudes described by distant poles.}
\end{figure}

\begin{figure}
\includegraphics[width=0.7\linewidth,angle=-90]{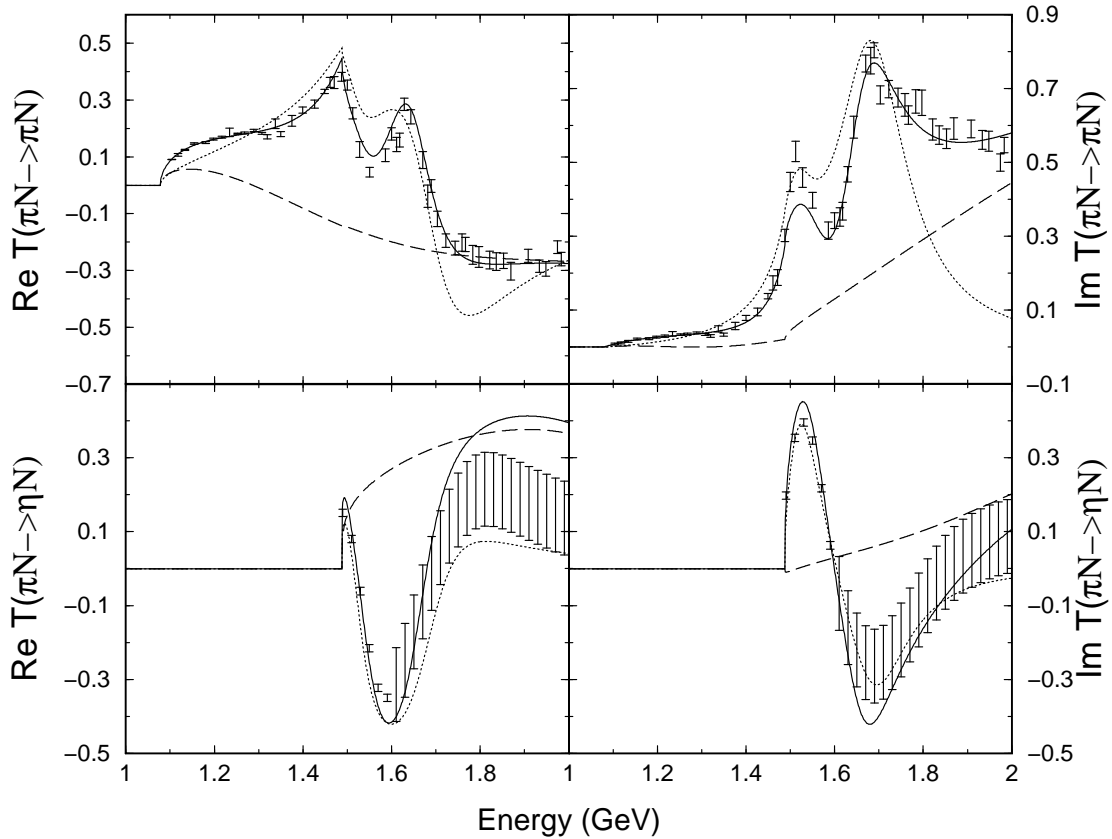}
\caption{\label{fig:k-distant} Caption as in
Fig.~\protect{\ref{fig:cmb-distant}}, except the resonant model used
here is the $K$-matrix model described in the text, with non-resonant
amplitudes described by distant poles.}
\end{figure}

\begin{figure}
\includegraphics[width=0.7\linewidth,angle=-90]{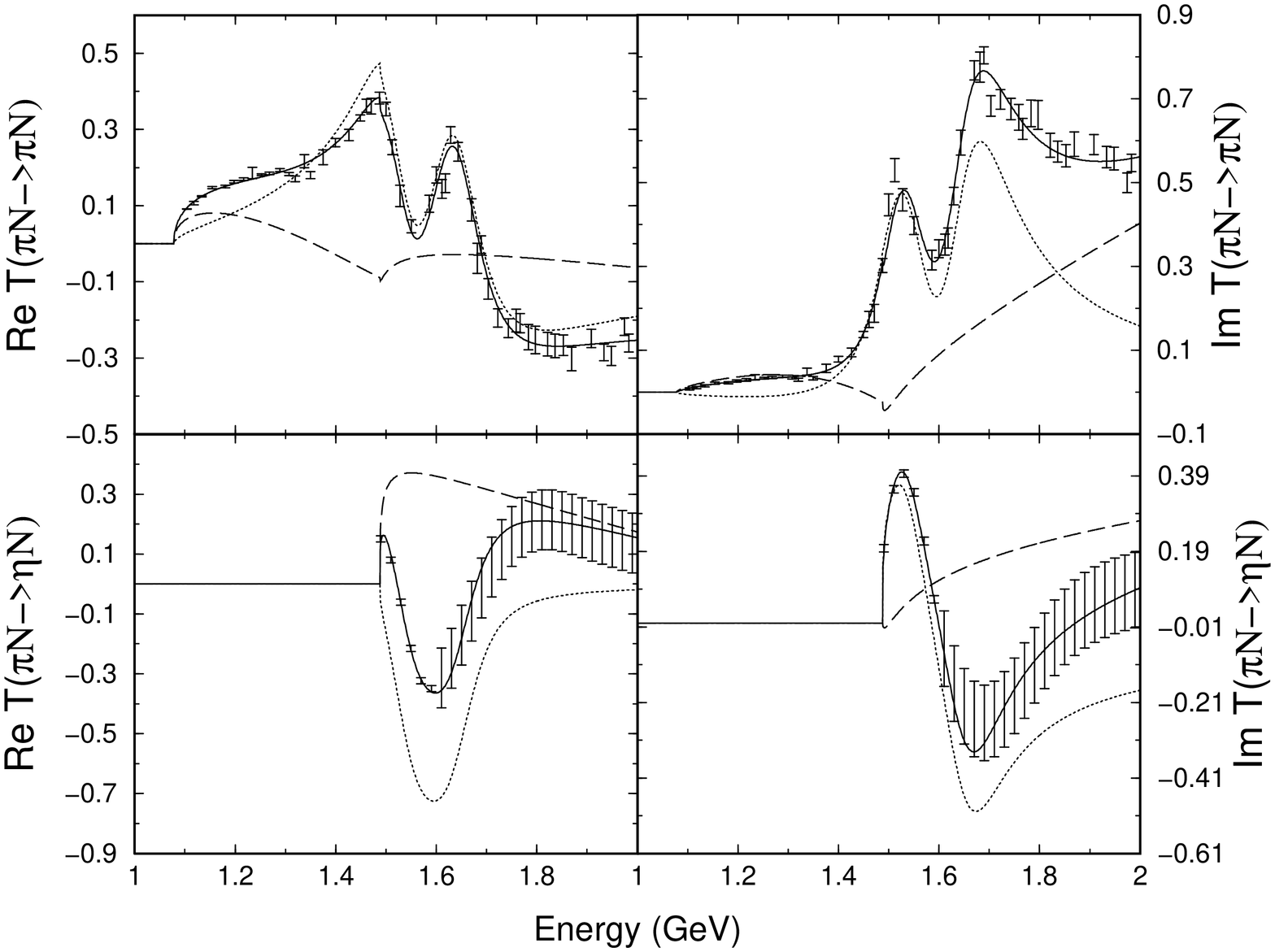}
\caption{\label{fig:bw-distant-nr} Caption as in
Fig.~\protect{\ref{fig:cmb-distant}}, except the resonant model used
here is the Breit-Wigner model with non-relativistic widths described in
the text, with non-resonant amplitudes described by distant
poles.}
\end{figure}

\begin{figure}
\includegraphics[width=0.7\linewidth,angle=-90]{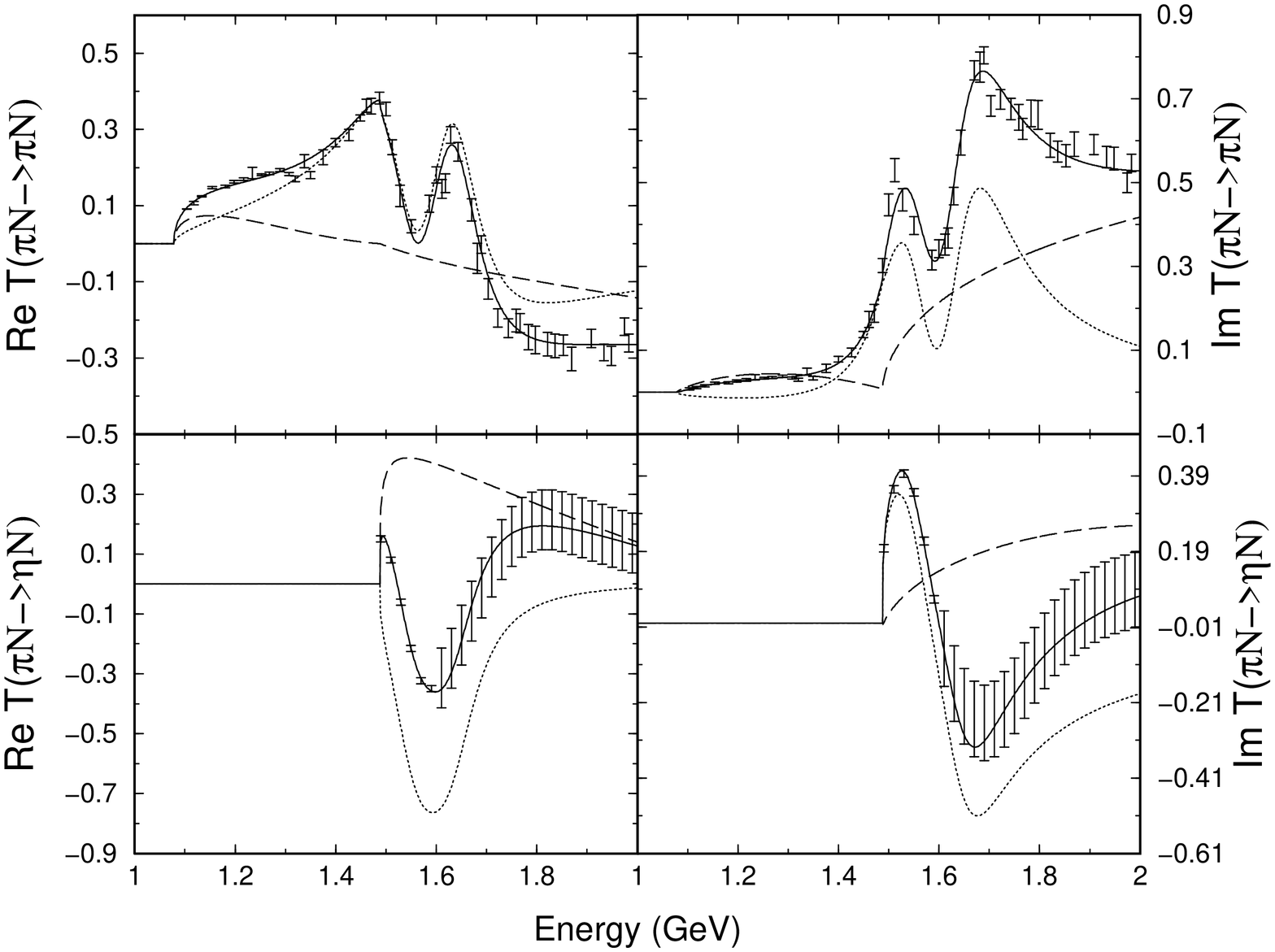}
\caption{\label{fig:bw-distant-rel} Caption as in
Fig.~\protect{\ref{fig:cmb-distant}}, except the resonant model used
here is the Breit-Wigner model with relativistic widths described in
the text, with non-resonant amplitudes described by distant
poles.}
\end{figure}

\begin{figure}
\includegraphics[width=0.7\linewidth,angle=-90]{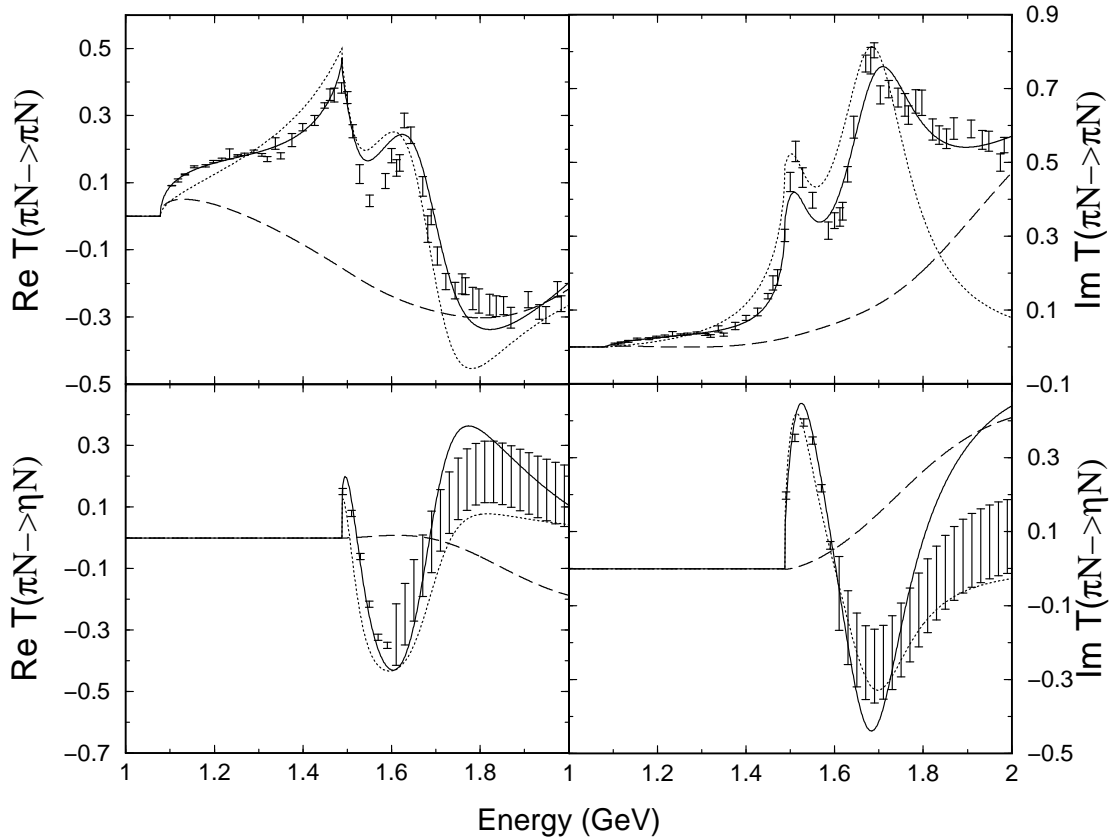}
\caption{\label{fig:cmb-poly} Caption as in
Fig.~\protect{\ref{fig:cmb-distant}}, except the resonant model used
here is the two-resonance, two channel CMB model described in the
text, with non-resonant amplitudes described by
polynomials.}
\end{figure}

\begin{figure}
\includegraphics[width=0.7\linewidth,angle=-90]{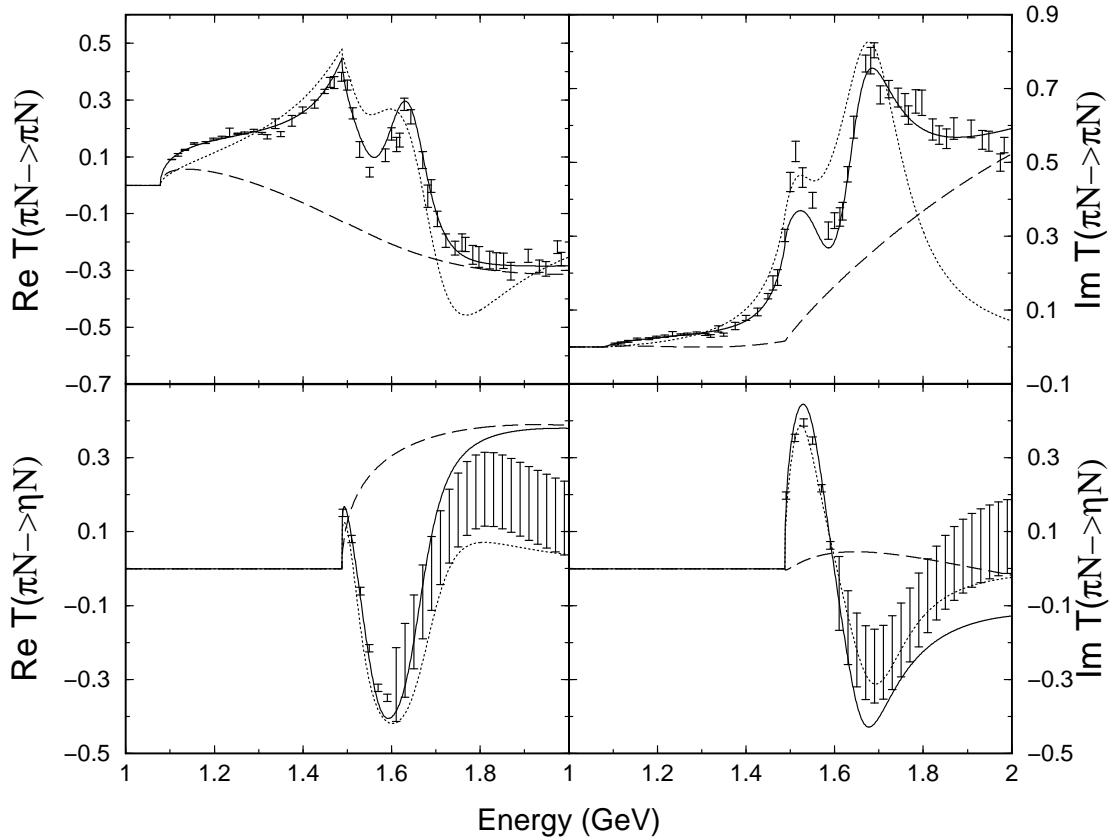}
\caption{\label{fig:k-poly} Caption as in
Fig.~\protect{\ref{fig:k-distant}}, except the resonant model used
here is the $K$-matrix model described in the text, with non-resonant
amplitudes described by polynomials.}
\end{figure}

\begin{figure}
\includegraphics[width=0.7\linewidth,angle=-90]{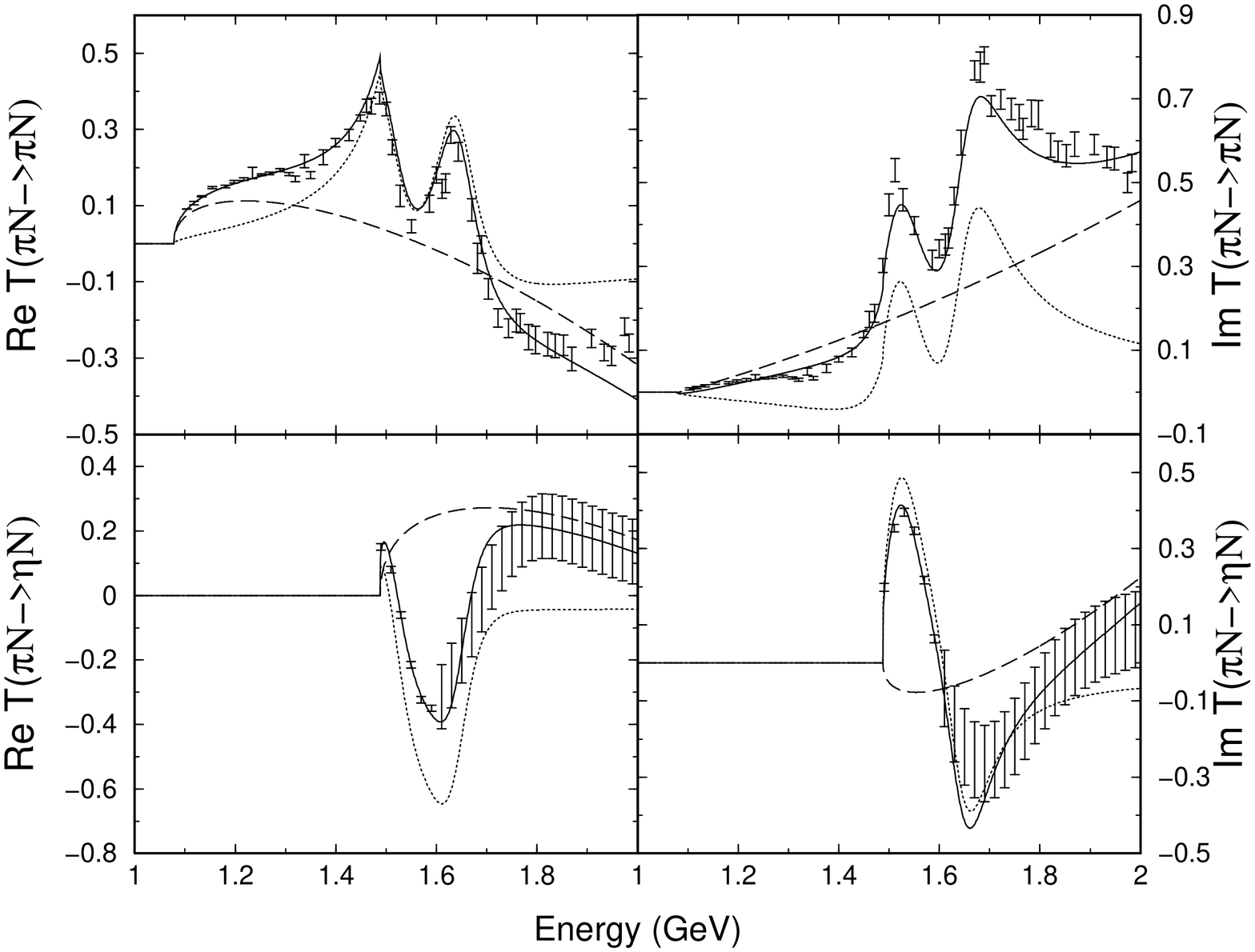}
\caption{\label{fig:bw-poly-nr} Caption as in
Fig.~\protect{\ref{fig:cmb-distant}}, except the resonant model used
here is the Breit-Wigner model with non-relativistic widths described in
the text, with non-resonant amplitudes described by
polynomials.}
\end{figure}

\begin{figure}
\includegraphics[width=0.7\linewidth,angle=-90]{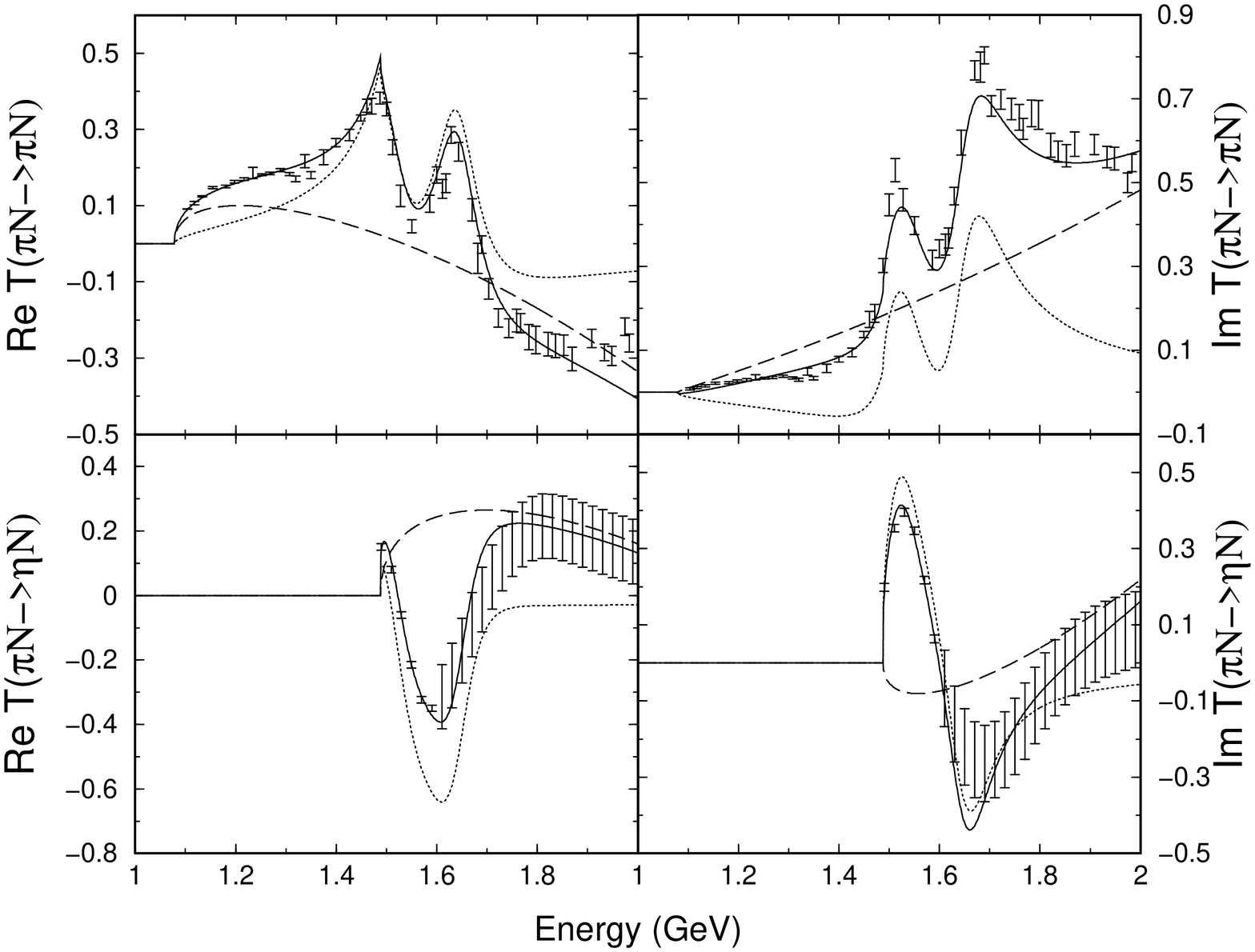}
\caption{\label{fig:bw-poly-rel} Caption as in
Fig.~\protect{\ref{fig:cmb-distant}}, except the resonant model used
here is the Breit-Wigner model with relativistic widths described in
the text, with non-resonant amplitudes described by
polynomials.}
\end{figure}

After this somewhat pedagogical introduction, the main results of this
work are now presented.  The fits to the $S_{11}$ partial-wave data
for the $T$-matrix elements $T_{\pi N\to \pi N}$ and $T_{\pi N\to \eta
N}$ for each resonant-non-resonant model are shown in
Figures~\ref{fig:cmb-distant} to~\ref{fig:bw-poly-rel}.  Unlike most
previous papers studying $N^*$ resonances, which show only the full
amplitude, the non-resonant amplitude (with all resonant couplings set
to zero) and the resonant amplitude (with all non-resonant couplings
set to zero) are also shown here. The effective Breit-Wigner
properties of the two resonances extracted from these fits are given
in Tables~\ref{tab-distant} and~\ref{tab-poly}. Errors in the first
four columns of results are determined from the fitting uncertainties only.

\begin{table}[t]
\begin{center}
\begin{tabular}{|c||c|c|c|c|c|c|c|}
\hline 
               model & CMB & $K$ & BW$_{\rm n.r.}$ & BW$_{\rm rel}$ 
& CMB  & CMB & PDG\\
               & & matrix & &
& all $\pi N$ & all $\pi N$, $\gamma N$ & \\
\hline \hline 
$S_{11}(1535)$ & & & & & & & \\
Mass (MeV)  & 1539$\pm$1 & 1533$\pm$1 & 1549$\pm$2 & 1558$\pm$1 & 1547$\pm 3$ 
& 1539$\pm$5 & 1520--1555(1535)\\
               Width (MeV) & 135$\pm$4  & 115$\pm$3  & 142$\pm$9  & 143$\pm$4  & 131$\pm$19  
& 122$\pm$20 & 100--200(150)\\
               $B_{\pi N}$ & 29$\pm$1\% & 34$\pm$1\% & 67$\pm$5\% & 67$\pm$1\% & 34$\pm$4\%  
& 39$\pm$5\% & 35--55\%\\
\hline \hline
$S_{11}(1650)$ & & & & & & & \\
Mass (MeV)  & 1682$\pm$1 & 1685$\pm$2 & 1648$\pm$5 & 1637$\pm$2 & 1690$\pm$12 
& 1684$\pm$15 & 1640--1680(1650)\\
               Width (MeV) & 144$\pm$3  & 190$\pm$5  & 147$\pm$10  & 145$\pm$4  & 227$\pm$40  
& 227$\pm$58 & 145--190(150)\\
               $B_{\pi N}$ & 80$\pm$1\% & 77$\pm$1\% & 74$\pm$5\% & 79$\pm$1\% & 75$\pm$3\%  
& 75$\pm$3 & 55--90\%\\
 \hline
               $\chi^2/N$  & 3.8  & 3.7  & 1.5 & 1.9  & 3.6         
& 5.6 & \\ 
 \hline
\end{tabular}
\caption{\label{tab-distant} Results for resonance parameters from
fits to the $T$ matrix elements for $\pi N\to \pi N$, $\eta N$ in the
$S_{11}$ partial wave, using the CMB, $K$-matrix, and Breit-Wigner
(BW) models described in the text. The third last column shows the
results of a fit including $\pi N\to \pi N$, $\eta N$ , $\rho N$, $\pi
\Delta$, $\sigma N$ and $\pi N^*$ partial wave
amplitudes~\protect{\cite{Vrana}}. The next to last column shows
results for a fit of the $\pi N$ data and $\gamma N \to \pi N$, $\eta
N$ data~\protect{\cite{Vranagam}}, and the last column shows the range
in the central values and estimated values from the
PDG~\protect{\cite{PDG}}. Non-resonant contributions to the $T$
matrix are described in terms of distant poles in all cases. Since the
primary results of this work come from a two-channel model, only the
branching fraction to $\pi N$ is given, since $B_{\eta N}=1-B_{\pi
N}$. The last row gives the $\chi^2$ per data point of each fit.}
\end{center}
\end{table}

\begin{table}[t]
\begin{center}
\begin{tabular}{|c|c||c|c|c|c|}
\hline 
& model        & CMB & $K$ matrix & BW$_{\rm n.r.}$ & BW$_{\rm rel}$ \\
\hline \hline 
$S_{11}(1535)$ & Mass (MeV)  & 1526$\pm$2  & 1533$\pm$1   & 1539$\pm$2   & 1538$\pm$2 \\
               & Width (MeV) & 112$\pm$6    & 119$\pm$3    & 130$\pm$6    & 130$\pm$6 \\
               & $B_{\pi N}$ & 30$\pm$2\%  & 33$\pm$1\%   & 39$\pm$1\%   & 38$\pm$1\%\\
\hline \hline
$S_{11}(1650)$ & Mass (MeV)  & 1688$\pm$2 & 1682$\pm$2   & 1648$\pm$2   & 1647$\pm$2 \\
               & Width (MeV) & 193$\pm$6  & 184$\pm$5    & 109$\pm$5    & 109$\pm$5 \\
               & $B_{\pi N}$ & 78$\pm$2\% & 75$\pm$1\%   & 51$\pm$1\%   & 51$\pm$1\%\\
 \hline
               & $\chi^2/N$  & 5.0   & 3.9   & 5.0    & 5.0 \\ 
 \hline
\end{tabular}
\caption{\label{tab-poly} Caption as in
Table~\protect{\ref{tab-distant}}, except non-resonant
contributions to the $T$ matrix are described in terms of
polynomials.}
\end{center}
\end{table}

A common criterion in fits is the value of $\chi^2$, given in the last
row of the tables. A comparison of the two-channel fits shows that the
lowest values of $\chi^2$ are attained using the Breit-Wigner models
with distant poles background, the more theoretically sophisticated
models are in between, and the Breit-Wigner models with polynomial
background have the highest $\chi^2$ values.  Breit-Wigner models
commonly have better fits because they are most often applied to study
single-channel reactions, but are much more likely to have additional
local minima close to the global minimum, reflecting the lack of
theoretical constraints. The extra parameters required to get a good
fit unfortunately obscure the physics results.  The extraction of
physically meaningful results for resonances depends more on the
quality of the theoretical constraints placed on the fit than on the
quality of the fit itself. In some cases the fit function is not as
sharp as the data, e.g., the CMB and $K$-matrix model Im($T_{\pi N \to
\pi N}$) amplitudes with polynomial background, although this is not
apparent from the total $\chi^2$, which is summed over all data
points.  All models are able to match the shape of the cusp at at
$\eta N$ threshold in Re($T_{\pi N \to \pi N}$), since this is a
property of interfering amplitudes and there are many possible
solutions.

In the figures more detail is shown than is customary.  For all
models, the amplitudes are separated into resonant and non-resonant
parts. The resonant amplitudes provide rapid energy fluctuations and a
rough match to the data, especially for $\pi N \to \eta N$.  The
non-resonant amplitudes are generally smooth.  However, analyticity
constraints require a cusp at $\eta N$ threshold, a feature of all
non-resonant amplitudes using distant poles background.  At first
glance, the fits all look similar.  More careful inspection reveals
differences in the detailed balance between resonant and
non-resonant amplitudes and in the channel coupling effects.  One of 
the most striking features is seen in the $\pi N \to \pi N$  
amplitude for the relativistic Breit-Wigner model with distant poles 
background, Fig.~\ref{fig:bw-distant-rel}.  The imaginary part 
of the non-resonant amplitude is much larger at $W \sim$ 1.7 GeV 
due to coupling to $\eta N$ than in the other models, a coupled 
channel effect.  The resonant part of the amplitude is then
smaller than in the other models and this model has a very small 
$B_{\eta N}$ for $S_{11}(1650)$.  

The resonance parameters extracted using the various models are
compared in Tables~\ref{tab-distant} and~\ref{tab-poly}. Variations
among the results for the resonance masses, the full width 
and the $\pi N$ branching fraction are all significant given the estimated
error bars of this study.  The estimated errors would have to be
much larger for the models to be in agreement.  For
reference, the result for the CMB model fitting two resonances with
all open channels is shown in Table~\ref{tab-distant}.  Note that the
errors quoted for the CMB results include systematic errors from model
uncertainties, and so are larger than the errors arising only from
fitting uncertainties in the results of the present work.  Many
resonance parameters found in the two-channel CMB model are within the
estimated errors of the full model.  The exception is the width of the
$S_{11}(1650)$, where the full CMB model value is larger than that of
any of the other eight models.  The full CMB model fit apparently has
a complicated interference between the second resonance and the
channels that are excluded in the truncated versions.  We also list the
recommended physical parameters for these states from the Particle
Data Group~\cite{PDG}.  Almost all results are within the conservative
estimated ranges they publish.  It is interesting that although
the results for the $S_{11}$(1535) full width have significant
variations using the different models studied here, the values are all
in the lower part of the PDG estimated range.  Similarly, the CMB and
$K$-matrix results for $B_{\pi N}$ are all in the lower part of the
PDG estimated range for that quantity.

Breit-Wigner models with relativistic and non-relativistic resonance
shapes are fitted separately and give almost identical results.
Despite having the best fits, they give results that vary the most
from the PDG values.  Compared to the other results, the Breit-Wigner
models have significantly lower mass, and smaller widths and $\pi N$
branching fractions for $S_{11}$(1650).  This is more true for the
fits using the polynomial background than for those using the distant
poles non-resonant amplitudes.  Since the resonant peak in Im($T$)
smoothly blends in with the non-resonant background, the $S_{11}$(1650)
width is very sensitive to how this background is treated.  Therefore,
it is not surprising that the biggest discrepancies arise in this
resonance property.  Interference with the overlapping lower energy 
state also has an important influence.

Comparisons between the CMB and $K$-matrix models are the most
interesting.  The full fit amplitudes in Figs.~\ref{fig:cmb-distant}
and \ref{fig:k-distant} are very similar.  They even miss the 
sharp structure in the $\pi N$ elastic amplitude at $W \sim$ 1.6
GeV in the same way.  At first glance the resonant and non-resonant
amplitudes are also very similar, but the small difference
in the shape of the imaginary elastic amplitude is the
primary source of the differences in the extracted full 
width in the two models.  The $S_{11}$(1535) full width from
the CMB model with distant poles background is larger than for 
the $K$-matrix model with distant poles, and vice versa for the 
$S_{11}$(1650) full width.  For the error bars we derive,
the difference in the full width is more significant than
the difference in $B_{\pi N}$.  This shows the interplay of the
interfering resonances.  On the other hand, the same two models
with polynomial background (where the CMB model no longer satisfies
analyticity) have very similar values for the
full width.  Interpretation of this result will be a key
component of the discussion.

\section{Discussion}

A variety of empirical models were used to fit partial wave amplitudes 
(the input `data') for a carefully
chosen problem, the $S_{11}$ partial wave and its two most important
channels, $\pi N$ and $\eta N$.  This partial wave is interesting
because of the physics interest in the $S_{11}$ resonances and the large
uncertainty in their properties as reported by the PDG~\cite{PDG}.  
By using identical input amplitudes and fitting strategies, we have 
made the {\it first objective comparison} of $N^*$ resonance extraction
models.  Four different resonance models (CMB, $K$-matrix, and 
Breit-Wigner with non-relativistic and relativistic widths), and two 
different empirical models for the
non-resonant amplitude (distant poles and polynomial) are employed.
These models are in regular use for the extraction of hadronic
properties.  It should be emphasized that the four resonance models
have almost identical amplitudes for isolated resonances.  The primary
differences among the models come from the way the dynamics of
resonance interference, multichannel effects, and non-resonant
amplitudes are treated.  Although resonance models of widely varying
quality are employed, the non-resonant models are both rather
empirical.  However, this partial wave in these channels has
non-resonant amplitudes which are comparatively small.  Since
the non-resonant amplitudes have a smooth energy dependence, their
influence on the extracted resonance parameters should be small.  
The main purpose of this work is to study the model dependence of the
extraction of $S_{11}$ properties (mass, width, and $\pi N$
branching fraction) in a case where overlapping resonances,
multichannel effects, and analyticity constraints are all expected to
be important.

The models used here can be put in order according to the theoretical
constraints employed.  The CMB model with distant poles background
satisfies multichannel unitarity and constraints from analyticity, and
handles resonance-resonance quantum mechanical interference well.  In
fact, the CMB model includes the most complete resonance propagation
effects of any of the existing models.  The $K$-matrix model used here
does not satisfy analyticity constraints, and leaves out re-scattering
dynamics present in the CMB model.  The Breit-Wigner models used here
satisfy essentially no theoretical constraints.  However, unitary
Breit-Wigner models~\cite{MAID} and $K$-matrix models that satisfy
analyticity constraints~\cite{Longacre} have been developed.  

The large range of properties for the two lowest energy $S_{11}$
resonances in the Review of Particle Properties~\cite{PDG} is also
found here.  This is evidence that much of the uncertainty in the PDG
estimates of $S_{11}$ properties comes from model dependence, since
the same input amplitudes are used in every fit.  No evidence is found
for a third $S_{11}$ state in the energy range studied.  Some of the
Breit-Wigner models have the best fits to the data, but this is due to
the flexibility of these models rather than an ability to describe the
underlying dynamics.  Arbitrary adjustments must be employed in order
to obtain good fits to the data.  The empirical phases between the
resonances provide a simple way to adjust the resonance-resonance
interference at the cost of obscuring the physics output.  As a
result, the physical properties of the $S_{11}$(1650) determined with
the Breit-Wigner models are very different than with the other models.

The models with the strongest theoretical constraints, the CMB and
$K$-matrix models with distant poles background, provide better
agreement with each other and with the CMB fits to a much larger set
of reactions~\cite{Vrana,Vranagam}.  One major result of the present
work is the differences between CMB and $K$-matrix models found in a
situation where their differences should be the largest.  The extent
of the disagreement cannot be simply stated.  For the error bars given
for the input amplitudes and those determined in our fits, the
difference between the $S_{11}$(1535) mass and full width for the CMB
and $K$-matrix models with distant poles is significant.  This is an
important measure of the difference in these two models.

Other considerations could contribute to these differences.  The
estimated errors quoted in the partial-wave amplitude fits have a
direct effect on the values quoted here.  It is possible that the
errors in the partial wave amplitudes are understated due to a lack of
understanding of the model dependence.  Although a more sophisticated
non-resonant amplitude could be required, the smoothness of these
amplitudes in all models argues against this.  Each model has problems
that are likely due to the truncated channel space employed.  The most
obvious problem is with the width of $S_{11}$(1650).  About 20\% of
the overall strength that was supposed to go to channels other than
$\pi N$ and $\eta N$ has to be included somewhere in the smaller
channel space.

In the context of the small but important test case chosen for this
study, the results of the CMB and $K$-matrix model fits are found to
have small, but potentially important differences.  Since the CMB
model is better constrained theoretically, the resonance properties
extracted using this model should be preferred when there is reason to
doubt other models.  A possible conclusion could be that this proves
that the constraints provided by analyticity, a superior treatment of
re-scattering, and an improved treatment of resonance interference are
important in this case. From the present limited comparison, such a
conclusion is likely premature. This issue requires further study,
particularly from the theoretical side.

We should note that many $N^*$ states will not be obscured by
strong threshold effects and will not have strong interference with
other states of the same angular momentum and parity.  The simplified
dynamics of the $K$-matrix model then give it the practical advantage of a
simpler and more stable path to a good fit to the partial-wave
amplitudes.  Proper use of this model in the analysis of $N^*$ data is
unlikely to give results that are badly incorrect. For resonances with
a larger number of open channels, the interaction of the non-resonant
and resonant amplitudes can be much more complicated and simplified
fitting can give erroneous results.

The primary result of this paper is that even in a small multi-channel
problem, dynamics are important.  Since Breit-Wigner models have very
few theoretical constraints, ad-hoc parameters are required to fit
real data such as those of the two-channel problem studied here.  
We therefore suggest that consideration of these issues be part
of any attempt to determine global recommendations for baryon 
resonance properties.

\section*{Acknowledgments}
This research is supported by the U.S. Department of Energy under
contracts DE-FG02-86ER40273 (AK, SC), and by the National Science
Foundation under grant PHY-0140116 (SD).


\begin{thebibliography}{9}

\bibitem{quarkmodels}
N.~Isgur and G.~Karl,
Phys.\ Rev.\ D {\bf 18}, 4187 (1978); 
S.~Capstick and N.~Isgur,
Phys.\ Rev.\ D {\bf 34}, 2809 (1986); 
L.~Y.~Glozman and D.~O.~Riska,
Phys.\ Rept.\  {\bf 268}, 263 (1996);
L.~Y.~Glozman, Z.~Papp and W.~Plessas,
Phys.\ Lett.\ B {\bf 381}, 311 (1996); 
see also the review by
S. Capstick and W. Roberts, Prog. Part. Nucl. Phys.  {\bf 45}, S241
(2000), and references therein.

\bibitem{chiral} T.~Inoue, E.~Oset and M.~J.~Vicente Vacas,
Phys.\ Rev.\ C {\bf 65}, 035204 (2002);
J.~Nieves and E.~Ruiz Arriola,
Phys.\ Rev.\ D {\bf 64}, 116008 (2001).

\bibitem{lattice1} M. G\"{o}ckeler, R. Horsley,
D. Pleiter, P.E.L. Rakow, G. Schierholz, C.M. Raynard, and D. Richards, 
Phys. Lett. {\bf B352}, 63 (2002).

\bibitem{lattice2} S. Dong, T. Draper, I. Horvath, F.X. Lee,
K.F. Liu, N. Mathur, and J.B. Zhang, LANL preprint hep-ph/0306199

\bibitem{Karl}
J.~Chizma and G.~Karl,
arXiv:hep-ph/0210126.

\bibitem{VPIpiN} Richard A. Arndt, Igor I. Strakovsky, Ron L. Workman, and 
Marcello M. Pavan, Phys. Rev. C{\bf 52}, 2120 (1995).

\bibitem{KH80} G. H\"{o}hler, F. Kaiser, R. Koch, and E. Pietarinen,
{\it Handbook of Pion-Nucleon scattering}, [Physics Data No. 12-1 (1979)].

\bibitem{KruMuk} B. Krusche, Nimai C. Mukhopadhyay, J.F. Zhang,
and M. Bennmerrouche, Phys. Lett. {\bf B397}, 171 (1997).

\bibitem{Vrana}
T.P. Vrana, S.A. Dytman and T.-S. H. Lee, Phys. Repts. {\bf 328}, 
181 (2000).

\bibitem{PDG}
K.~Hagiwara {\it et al.}  [Particle Data Group Collaboration],
Phys.\ Rev.\ D {\bf 66}, 010001 (2002).

\bibitem{Baker}
R.D. Baker et al. Nucl. Phys. {\bf B156}, 93 (1979).

\bibitem{MAID}
D. Drechsel, O. Hanstein, S.S. Kamalov, and
L. Tiator, Nucl. Phys. {\bf 645}, 145, (1999).

\bibitem{Penner}
G. Penner and U. Mosel, Phys. Rev. C{\bf 66}, 055211 (2002);

\bibitem{Green} A.M. Green and S. Wycech, Phys. Rev. C{\bf 55}, R2167 (1997).

\bibitem{Batinic}
M.~Batinic, I.~Slaus, A.~Svarc, and B.M.K. Nefkens,
Phys.\ Rev.\ C {\bf 51}, 2310 (1995).

\bibitem{Cutkosky} R.E. Cutkosky, C.P. Forsyth, R.E. Hendrick, and R.L. Kelly,
Phys. Rev. D{\bf 20}, 2839 (1979).

\bibitem{Manley}
D.~M.~Manley,
Int.\ J.\ Mod.\ Phys.\ A {\bf 18}, 441 (2003).

\bibitem{Longacre}
R.S. Longacre, T. Lasinski, A.H. Rosenfeld, G. Smadja, R.J. Cashmore, 
and David W.G.S. Leith, Phys.Rev. D{\bf 17} 1795 (1978). 

\bibitem{Moorhouse} R.G. Moorhouse, H. Oberlack, and A.H. Rosenfeld,
Phys. Rev. D{\bf 9}, 1 (1974).

\bibitem{Chung} 
S.~U.~Chung, J.~Brose, R.~Hackmann, E.~Klempt, S.~Spanier and C.~Strassburger,
Annalen Phys.\  {\bf 4}, 404 (1995).

\bibitem{Vranagam} T. Vrana, S.A. Dytman, and T.-S.H. Lee, to 
be published.

\end{thebibliography}
\end{document}